\documentclass[aps,prl,floatfix,twocolumn,superscriptaddress]{revtex4-1}
\usepackage{amsmath,amssymb}
\usepackage{graphicx}
\usepackage{epsfig}
\usepackage{psfrag}
\usepackage[usenames]{color}
\usepackage{xcolor}
\usepackage{hyperref}
\usepackage{soul,color}
\usepackage{physics}
\usepackage{tabu}
\usepackage{multirow}
\usepackage{ulem}
\hypersetup{colorlinks=true,citecolor={blue},linkcolor={blue},urlcolor={blue}}

\begin{document}
\title{Valley Acoustoelectric Effect}

\author{A.~V.~Kalameitsev}
\affiliation{Rzhanov Institute of Semiconductor Physics, Siberian Branch, Russian Academy of Sciences, Novosibirsk, 630090 Russia}

\author{V.~M.~Kovalev}
\affiliation{Rzhanov Institute of Semiconductor Physics, Siberian Branch, Russian Academy of Sciences, Novosibirsk, 630090 Russia}
\affiliation{Novosibirsk State Technical University, Novosibirsk, 630072 Russia}
\date{\today}

\author{I.~G.~Savenko}
\email[Corresponding author: ]{ivan.g.savenko@gmail.com}
\affiliation{Center for Theoretical Physics of Complex Systems, Institute for Basic Science (IBS), Daejeon 34126, Korea}
\affiliation{Rzhanov Institute of Semiconductor Physics, Siberian Branch, Russian Academy of Sciences, Novosibirsk, 630090 Russia}

\begin{abstract}
We report on novel valley acoustoelectric effect, which can arise in a 2D material, like a transition metal dichalcogenide monolayer, residing on a piezoelectric substrate. 
The essence of this effect lies in the emergence of a drag electric current (and a spin current) due to a propagating
surface acoustic wave.
This current consists of three contributions, one independent of the valley index and proportional to the acoustic wave vector, the other arising due to the trigonal warping of the electron dispersion, and the third one is due to the Berry phase, which Bloch electrons acquire traveling along the crystal.
As a result, there appear components of the  current orthogonal to the acoustic wave vector. 
Further, we build an angular pattern, encompassing nontrivial topological properties of the acoustoelectric current, and suggest a way to run and measure the conventional diffusive, warping, and acoustoelectric valley Hall currents independently.
We develop a theory, which opens a way to manipulate valley transport by acoustic methods, expanding the applicability of valleytronic effects on acousto-electronic devices.
\end{abstract}


\maketitle


Two-dimensional materials (2D materials), such as transition metal dichalcogenides~\cite{RefRadisavljevic,  RefSundaram, xu2014spin}, possess symmetry properties similar to graphene~\cite{Geim}.
Their primary feature is that the valleys K and K$^\prime$ in the Brillouin zone connect by the time reversal symmetry.
Consequently, the chiralities of the K and K$^\prime$ bands turn out opposite, and in addition to conventional momentum and spin of the two-dimensional electron gas (2DEG), 2D materials acquire an additional \textit{valley index} degree of freedom.
Moreover, their spectra manifest large gaps in the optical range~\cite{xiao2012coupled}, epitomizing various valley-resolved phenomena~\cite{RefMak, RefUbrig}.

Exposed to external strong electromagnetic fields resulting in a dynamical gap opening~\cite{OurNJP, ShelykhGap, Machlin, BlochOsc}, 2D materials can exhibit such fascinating phenomena as dissipationless transport of 2DEG~\cite{KibisPRL} and the photon drag effect~\cite{Wieck, Glazov, Entin, RefRPQ}.
%
Moreover, stemming from the essential spatial inversion symmetry breaking, there might occur transport phenomena described by a third-order conductivity tensor~\cite{Glazov, RefBasov}, finite in noncentrosymmetric materials.
For example, in the photovoltaic effect~\cite{OurRecPRB}, the conductivity tensor $\chi$ couples components of the photoinduced current $j_\alpha$ with the components of the external electric field $E_\beta$:
$j_\alpha=\chi_{\alpha\beta\gamma}E_\beta E_\gamma$, where $\alpha,~\beta,~\gamma=x,~y,~z$.

The conventional photovoltaic effect originates from an asymmetry of the interaction potential or the 
Bloch wave function~\cite{belinicher}.
In 2D materials, there can also appear an unconventional mechanism of this effect, which is due to the trigonal warping of the valley spectrum, resulting in the asymmetry of the interband optical transitions~\cite{Shan}.
Beside the valley and spin currents~\cite{MEGT, Hongyi, ZhangScience}, trigonal warping also manifests itself in the second-harmonic generation phenomena~\cite{GT}, spin-resolved measurements of the photoluminescence from the sample, and alignment of the photoexcited carriers in gapless materials~\cite{Portnoi1}.

Only few phenomena distinguish Bloch electrons from free charges, and one of them is the Berry effect, which, in particular, influences the carriers of charge subject to a mechanical force $e\mathbf{E}$, where $\mathbf{E}$ is an external electric field and $e$ is the elementary charge. 
It happens since the group velocity of a Bloch electron acquires an additional anomalous term $e\mathbf{E}\cross\mathbf{\Omega}_\mathbf{k}$, where $\mathbf{k}$ is the momentum of the particle and $\mathbf{\Omega}_\mathbf{k}$ is the Berry curvature~\cite{RefChang, RefSundaram}. 
In the framework of the linear response theory, the matrix of the velocity operator acquires nonzero off-diagonal linear in field elements, thus mixing different bands.

The concept of the Berry phase~\cite{RefBerry} underlies and unifies diverse aspects of solid-state physics, drastically affecting the transport of particles and resulting in such intriguing phenomena as the anomalous~\cite{RefNagaosa} and quantum~\cite{RefThouless} Hall effects, emergence of topological and superconducting phases~\cite{RefHasan}, charge pumping~\cite{RefNiu}, anomalous thermoelectric transport~\cite{RefXiao}, among other~\cite{RefXiaoRMP}. 
By and large, electrons in a crystal behave similarly to free particles with just the free electron mass replaced by an effective one due to the formation of energy bands. 
A nontrivial Berry phase also reveals itself in Dirac materials like graphene~\cite{RefZhang}, and from very recently it comes in play in other two-dimensional (2D) materials~\cite{RefYou}, possessing similar symmetry properties.

\begin{figure*}[!t]
\includegraphics[width=0.99\textwidth]{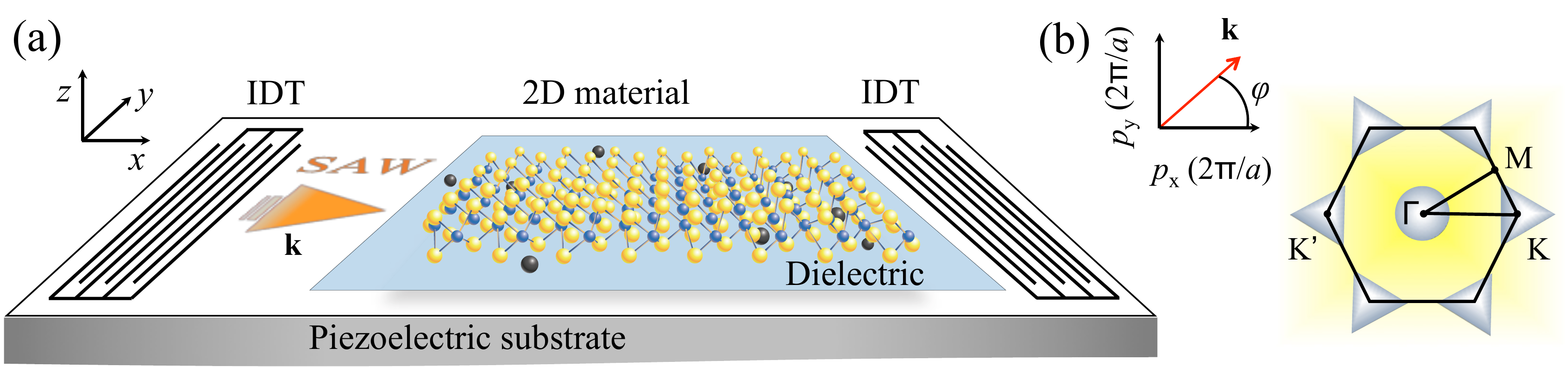}
\caption{System schematic.
(a)
2D material (MoS$_2$), exposed to a surface acoustic wave (SAW) with the wave vector $\mathbf{k}$. The sample lies on a layer of dielectric on a piezoelectric substrate. Two interdigital transducers (IDTs) generate and absorb the SAWs.
(b)
The first Brillouin zone of MoS$_2$ with the schematic illustration of warping.}
\label{Fig1}
\end{figure*}

Instead of light, surface acoustic waves (SAWs) can be employed to probe or alter the physical properties of electron (and other) gases in low-dimensional systems~\cite{RefWixforth, RefWillet}.
It is relatively simple to launch SAWs in piezoelectric heterostructures. That is why SAWs are frequently used in engineering and scientific applications, forming the basis of acoustoelectronics.
The appearance of new 2D materials stimulates the studies of SAWs, interacting with electrons in graphene monolayers~\cite{graphene1, graphene2}, surfaces of 
topological insulators~\cite{top1}, and thin films~\cite{top2}. 
Recently, 
there have been suggested SAW spectroscopy methods to study 2D dipolar exciton gases in normal and Bose-condensed phases~\cite{reviewkovalev1, reviewkovalev2}, including the acoustic drag effect~\cite{dragkovalev1}.
%
Experimentally, one can either (i) measure the absorption of sound by a 2DEG, or (ii) observe renormalization of the SAW velocity in heterostructures exposed to strong magnetic fields~\cite{RefEsslingen} due to their interaction with the carriers of charge, or (iii) study the acoustoelectric (AE) effect. The latter consists in the emergence of stationary electric currents when a SAW draggs the carriers of charge via the momentum transfer to the 2DEG~\cite{Parmenter}.

In this Letter, we demonstrate that in multivalley 2D materials, there take place an unconventional AE effect  and an AE Valley Hall effect (AVHE). 
We consider a transition metal dichalcogenide monolayer, taking MoS$_2$ as an example, and show that the trigonal valley warping gives an additional component of the AE current with peculiar properties, characteristic of 2D materials (we will call this component the \textit{warping current}). 
Furthermore, the Berry effect gives an unconventional acoustic drag Hall current. 
It is known~\cite{RefXiaoRMP} that if a TMD monolayer is exposed to an in-plane static electric field, the Berry curvature allows for the appearance of the valley Hall effect, when the current flows in the direction transverse to the static in-plane electric field. 
If we take a cw instead of the static field, the stationary valley Hall current is absent since the time-averaged force acting on electrons vanishes. 
However, a nonzero force appears in the second order with respect to the cw electric field. 
Here we consider the case when such a force is due to the piezoelectric field of the surface acoustic waves, traveling along the surface of the piezoelectric substrate.
We show that the joint influence of this force and the Berry phase allow for the AVHE.
These currents couple with the piezoelectric field of an external acoustic wave via the third-order conductivity tensor, as in the photovoltaic effect mentioned above, forming various fascinating propagation patterns.
Moreover, the SAWs aspire to separate particles with opposite spins, resulting in a spin current.

\begin{figure*}[!t]
\includegraphics[width=0.99\textwidth]{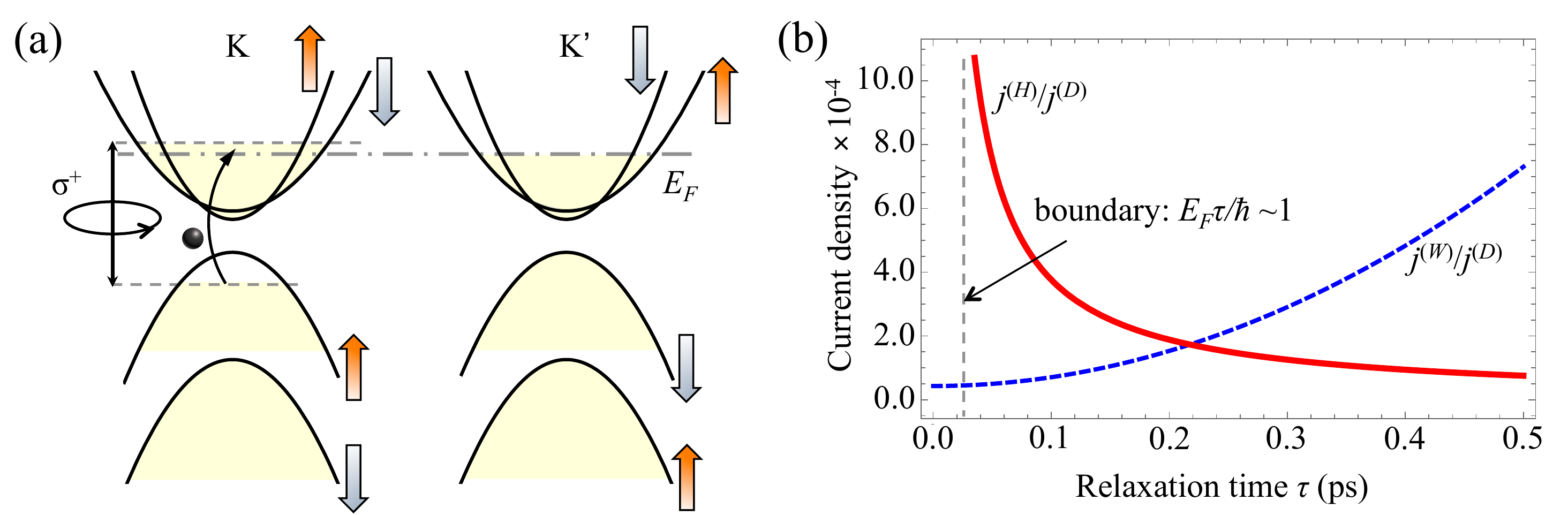}
\caption{(a) The band structures of MoS$_2$ with account of the optically induced imbalance of the valley populations. 
Yellow shaded regions indicate the filled states. 
The arrows signify the directions of spin in each valley. 
(b) Relative magnitudes of the AE warping (dashed blue) and valley Hall (solid red) components of the current density as functions of the relaxation time at $n=5\times10^{12}$~cm$^{-2}$.}
\label{Fig2}
\end{figure*}

Let us consider a layer of MoS$_2$, separated from a piezoelectric substrate by a dielectric layer (Fig.~\ref{Fig1}). A Bleustein-Gulyaev SAW with the wave vector $\textbf{k}$  travels along the interface and creates a piezoelectric field having both the out-of-plane and in-plane components. The latter is $\textbf{E}||\textbf{k}$ and it acts on the 2DEG.
This field drags the carriers of charge in MoS$_2$, resulting in the AE current. We assume that the monolayer is n-doped. %
Furthermore, the conduction band in each of the valleys is split by spin due to the spin-orbit interaction (SOI), as is shown in Fig.~\ref{Fig2}(a); the strength of the SOI for MoS$_2$ is of the order of $3$ meV~\cite{RefFalkoPar}.

The group velocity describing the quasiclassical dynamics of a Bloch electron in the absence of an external magnetic field reads
\begin{gather}\label{EQ.1}
\dot{\textbf{r}}=\mathbf{v}-\dot{\textbf{p}}\times {\bf \Omega}_\textbf{p},
\end{gather} 
where $\mathbf{v}=\partial\varepsilon_{\textbf{p}}/\partial\textbf{p}$, $\varepsilon_\textbf{p}=\textbf{p}^2/2m+w_\textbf{p}$ is the electron dispersion in a given valley with account for its warping $w_\textbf{p}=\eta C(p_x^3-3p_xp_y^2)$,  $\eta=\pm 1$ is a valley index, $C$ is a warping strength, and $\dot{\textbf{p}}=e\tilde{\textbf{E}}$ with $e<0$ the electron charge, and $\tilde{\textbf{E}}(\textbf{r},t)=\tilde{\textbf{E}}e^{i\textbf{kr}-i\omega t}/2 +\textrm{c.c.}$ is the overall electric field, including the piezoelectric ${\bf E}({\bf r},t)$ and induced $\textbf{E}^i(\textbf{r},t)$ contributions.
The origin of the induced electric field $\textbf{E}^i(\textbf{r},t)$ is the fluctuations of the electron density.
The Berry curvature reads ${\bf \Omega}_\textbf{p}=\partial_\textbf{p}\times\langle u|i\partial_\textbf{p}|u\rangle$ and $|u\rangle$ is a periodic amplitude of the Bloch wave function. 

%
%
%

To describe the electron transport, we will use the Boltzmann transport equation~\cite{Chaplik} %
\begin{gather}\label{eq1}
\frac{\partial f}{\partial t}+\dot{\textbf{p}}\cdot\frac{\partial f}{\partial \textbf{p}}+\dot{\textbf{r}}\cdot\frac{\partial f}{\partial \textbf{r}}=I\{f\},
\end{gather}
where $f$ is the electron distribution function, $I\{f\}$ is the collision integral.
For the collision integral we use the model of a single--$\tau$ approximation~\cite{Kittel}, which does not depend on energy: $I\{f\}=-(f-\langle f\rangle)/\tau$.
Here $\langle f\rangle$ is the locally-equilibrium distribution function in the reference frame moving with the SAW. It depends on the local electron density $N(\textbf{r},t)$ via the chemical potential $\mu=\mu(N)$. 
Furthermore, we expand the density in series: $N(\textbf{r},t)=n+n_1(\textbf{r},t)+n_2(\textbf{r},t)+O(n_3)$, where $n$ is the unperturbed electron density and $n_i$ are the corrections to the density fluctuations.
We expect that the AE current should appear as the second-order response to the external piezoelectric field. 
Thus we expand the distribution function: $f=f_0+f_1+f_2+O(f_3)$, where $f_0=(\exp\{[\varepsilon_\mathbf{p}-\mu(n)]/T\}+1)^{-1}$ is the equilibrium electron distribution, which depends on the electron momentum $\textbf{p}$ only. 
We also expand $\langle f\rangle$: $\langle f\rangle=f_0+(n_1+n_2+...)\partial_n f_0+(n_1+n_2+...)^2\partial^2 f_0/\partial n^2/2$.
The induced electric field obeys the Maxwell's equation
$\textmd{div}\, \textbf{D}^i=4\pi\rho$,
where $\mathbf{D}^i=\epsilon(z)\mathbf{E}^i$, $\epsilon(z)$ is the dielectric function, and the charge density reads  $\rho=e(N({\bf r},t)-n)\delta(z)$. The solution is
$\textbf{E}^i=-4\pi ie\textbf{k}(N-n)_{{\bf k},\omega}/[(\epsilon+1)k]$,
where $\epsilon$ is the dielectric constant of the substrate.




\textit{{Results and discussion.---}} 
Let us, first, assume that the  warping and the Berry phase are absent.
In this case the drag of electrons is valley-independent. 
For a SAW traveling with the momentum $\textbf{k}$ and in the long-wavelength limit ($\omega\tau,~\textbf{k}\cdot\mathbf{v}\tau\ll1$) the drift current is negligibly small, whereas for a degenerate electron gas at zero temperature (which, in particular, gives $\partial\mu/\partial n=\pi/m$), we find the  diffusive current (see Supplemental Material~\cite{SM})
\begin{eqnarray}\label{eq14}
\mathbf{j}^{(D)}=
\frac{e\tau}{2m}\frac{\mathbf{k}\sigma}{\omega}\frac{|E_0|^2}{1+\left(kv^*/\omega+Dk^2/\omega\right)^2},
\end{eqnarray}
%
%
%
where $\sigma=e^2n\tau/m$ is a 2D static Drude conductivity with the dimensionality of velocity, $D=v_F^2\tau/2$ is a diffusion coefficient, $v_F$ is the Fermi velocity,  $v_*=4\pi\sigma/(\epsilon+1)$ is the velocity of charge spreading in 2D systems, and $E_0$ is the piezoelectric field amplitude. 
We want to note, that Eq.~\eqref{eq14} at $Dk^2/kv_*\ll1$ coincides with the formula of AE current, reported in Ref.~\onlinecite{Falko1993}.
\begin{figure*}[!t]
\includegraphics[width=0.99\textwidth]{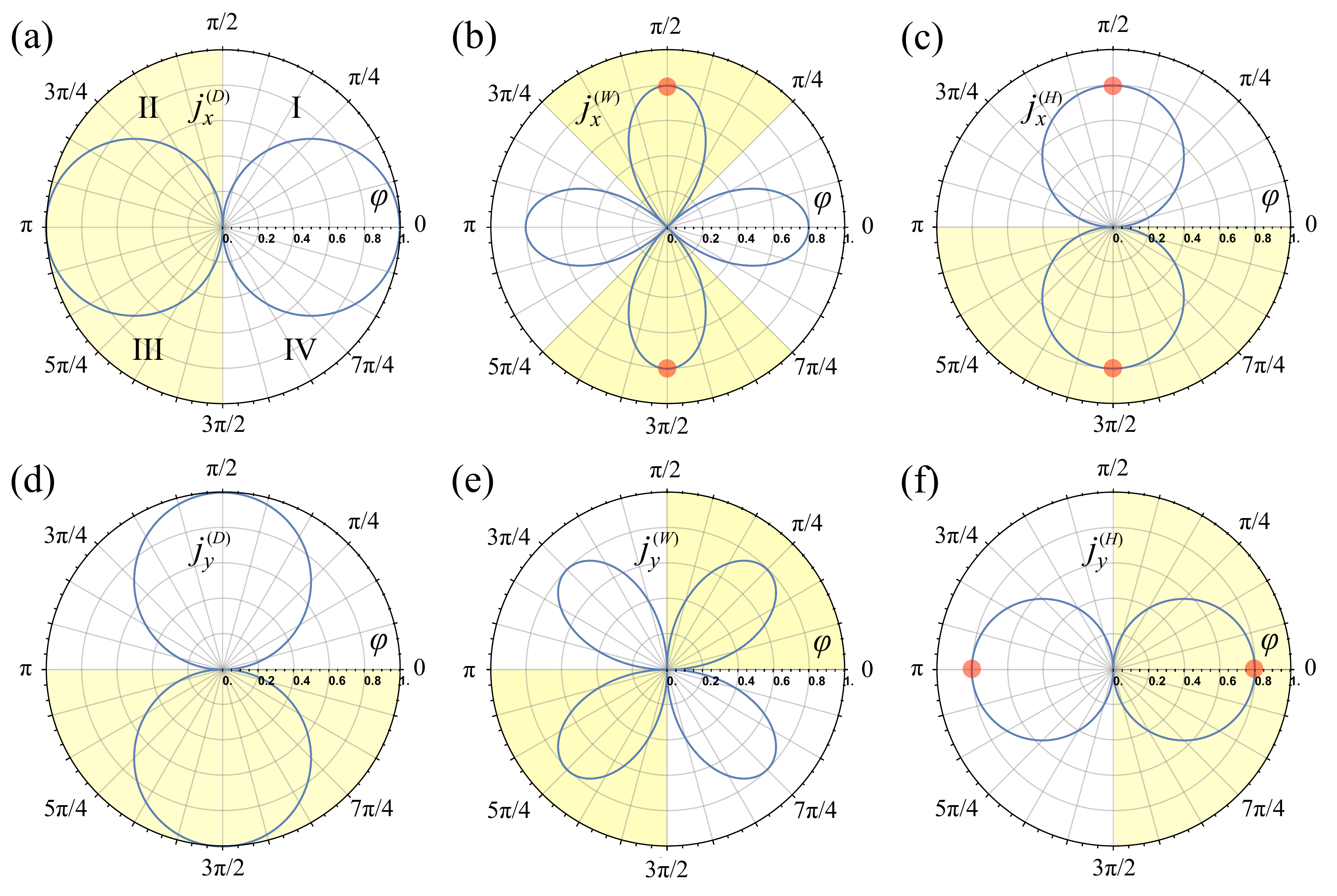}
\caption{Angular patterns of the $x$ and $y$ components of the diffusive (a,d), warping (b,e) and Hall (c,f) current density in arbitrary units.
The warping and Hall current petals are depicted smaller than the conventional diffusive current petal to remind that their magnitudes are smaller. 
Yellow shading marks the areas of negative current (directed opposite to x- or y-axis).
Red dots manifest the angles ($\pi/2$, $3\pi/2$, $0$ and $\pi$), at which only the unconventional current flows along the $x$- or $y$-direction.}
\label{Fig3}
\end{figure*}
%
%
%


%
%
%
%
%

Now we switch the trigonal warping and the Berry phase on. 
After derivations~\cite{SM} we find the components of the current density.
Let us compare the AE diffusive, warping~\cite{Hongyi} and Hall currents.
They can be written in the uniform way:
\begin{eqnarray}
\label{EqDifCurrent}
\textbf{j}^{(D)}&=&
\frac{e\sigma k}{2\omega}\frac{\tau}{m}
\frac{\textbf{n}}{1+\left(\sigma/\sigma_*\right)^2(1+ka)^2}E_0^2,\\
\label{EqWarpingCurrent}
\textbf{j}^{(W)}&=&e\sigma m\tau\frac{\nabla_{\textbf{n}}C(\textbf{n})\left[1+\left(\sigma/\sigma_*\right)^2(ka)^2\right]}{1+\left(\sigma/\sigma_*\right)^2(1+ka)^2}E_0^2,\\
\label{EqHallCurrent}
\textbf{j}^{(H)}&=&\frac{e\sigma k}{2\omega}\frac{[\textbf{n}\times{\bf \Omega}_0]}{1+\left(\sigma/\sigma_*\right)^2(1+ka)^2}E_0^2,
\end{eqnarray}
where we choose the coordinate axes as in Fig.~\ref{Fig1}, then $\textbf{n}=\textbf{k}/k$, $C(\textbf{n})=\eta C(n_x^3-3n_xn_y^2)$, $\sigma_*=(\epsilon+1)s/4\pi$ and $a=(\epsilon+1)\hbar^2/(4me^2)$. 
The Berry curvature has an out-of-plane component ${\bf \Omega}_0=(0,0,\hbar\eta/m\Delta)$, where $\Delta$ is the band gap of the TMD monolayer. 
Due to the usual smallness of the warping, we will disregard its contribution to the static conductivity and the charge spreading, when describing the screening effect.

From Eqs.~(\ref{EqWarpingCurrent}) and~(\ref{EqHallCurrent}) we see that the net valley AE currents summed over the valley indexes $\eta=\pm 1$ are zero, due to the time-reversal symmetry. Hence, it should be broken to detect the valley currents. 
One of the possibilities to do it is to expose the sample to a circularly-polarized light with the frequency close to the band gap since the optical selection rules in 2D materials depend on $\eta$, see Fig.~\ref{Fig2}(a). 
Then one of the valleys will be dominantly populated. 
The difference in the particle densities, $\delta n=n(\eta=1)-n(\eta=-1)\neq 0$, results in a nonzero net valley current.

Let us now estimate relative magnitudes of different contributions to the AE current. 
For realistic parameters $\sigma/\sigma_*\gg1$ and $ka\ll1$ we find
\begin{gather}\label{EstCurrents1}
j^{(D)}\approx\frac{(\sigma_*E_0)^2}{ens}\approx 20\,\mu \textrm{A/cm},
\end{gather}
where we used the sound velocity $s=3.5\cdot10^5$~cm/s for LiNbO$_3$ piezoelectric substrate, acoustic wave piezoelectric potential amplitude $\varphi_\textrm{SAW}=50$~mV (thus $E_0=k\varphi_\textrm{SAW}$), and other parameters read $n=5\cdot10^{12}\,cm^{-2}$, $\tau=2\cdot10^{-13}\,s$, $\omega=10^{10}\,s^{-1}$.

The AVHE contribution stemming from the Berry-phase effect relates to the diffusive current as
\begin{gather}\label{EstCurrents2}
j^{(H)}/j^{(D)}=\frac{\delta n}{n}\frac{\hbar}{\tau\Delta},
\end{gather}
whereas for the warping current,
\begin{gather}\label{EstCurrents3}
j^{(W)}/j^{(D)}=\frac{\delta n}{n}\left(\frac{C_ems^2}{\hbar^3}\right)\left[1+\left(\sigma/\sigma_*\right)^2(ka)^2\right].
\end{gather}
Now taking $C_e = -3.49$ eV$\cdot {\AA}^3$ and $m=0.44m_0$ (for MoS$_2$), $\delta n/n=0.1$, we find the estimations $j^{(H)}\sim 4$~nA/cm and $j^{(W)}\sim 3$~nA/cm.

Note that the warping and Hall currents depend differently on the relaxation time [see Fig.~\ref{Fig2}(b)]. 
The $\tau-$ dependence of warping current (\ref{EstCurrents3}) is determined by the $\sigma^2$ term, whereas the Hall current~(\ref{EstCurrents2}) grows with the decrease of $\tau$. However, the system imposes the lower boundary of $\tau$ dictated by the condition $E_F\tau/\hbar\gg1$, which implies $\tau\approx 0.2$~ps at $n=5\times 10^{12}$~cm$^{-2}$.
The upper boundary is determined by the applicability of the diffusive limit.

The AE Hall and warping currents might have comparable magnitudes in some parameter range but they possess different topological properties.
Indeed, the  current densities~\eqref{EqDifCurrent},~\eqref{EqWarpingCurrent}, and~\eqref{EqHallCurrent} have the forms $\mathbf{j}^{(D)}=j_0^{(D)}(\cos\varphi,\sin\varphi)$, $\mathbf{j}^{(W)}=j_0^{(W)}(\cos2\varphi,-\sin2\varphi)$, and $\mathbf{j}^{(H)}=j_0^{(H)}(\sin\varphi,-\cos\varphi)$, respectively (see Fig.~\ref{Fig3}).
The magnitudes of both the diffusive and Hall current densities are proportional to the SAW wave vector. 
Thus they behave like cosine or sine of the angle, describing the direction of propagation of the SAW. 
In the mean time, the warping-related current originates from the warping of the electron dispersion in the valleys, which behaves as $\cos 3\varphi$ due to the C$_{3h}$ symmetry group. 
Furthermore, the electron velocity, being the derivative of the energy with respect to the electron momentum, gives the $\cos 2\varphi$ and $\sin2\varphi$ behavior of the warping-related current density.

Let us consider an acoustic wave propagating along the $y-$direction, which corresponds to $\varphi=\pi/2$ or $\varphi=3\pi/2$. 
Then the $x-$component of the diffusive current vanishes, while the $x-$components of the AE Hall and warping currents are nonzero [see Fig.~\ref{Fig3}(b,c)]. 
If $\varphi=\pi/2$, the two currents have opposite direction and partially compensate each other, whereas if $\varphi=3\pi/2$, the currents sum up. Obviously, it allows to distinguish between their contributions.
If the acoustic wave propagates along the $x-$direction, then $\phi=0$ or $\phi=\pi$ and only the $y-$component of the Hall current is nonzero [see Fig.~\ref{Fig3}(f)], while the $y-$components of the diffusive and warping currents vanish. 

We want to note, that the SOI for the conduction band, being small in comparison with typical optical frequencies, is usually disregarded in optically-induced transport effects. 
We have estimated the relative contributions of the AE warping and Hall currents and  shown that they can have comparable magnitudes for the electron densities of the order $n=5\cdot 10^{12}$~cm$^{-2}$. 
For such density, the Fermi energy lies deep in the conduction band, exceeding the  SO splitting energy (for the MoS$_2$ it amounts to 3~meV). 
In these condictions the spin current is negligibly small. 
A possible way to observe the AE spin effect is to have a p-doped layer with Fermi energy lying in the valence band between the SO-split hole subbands with the splitting $\sim 400$~meV for transition metal dichalcogenide monolayers, which exceeds by orders the SAW frequencies. 
Obviously, the theory developed in this work is directly applicable to the p-doped TMDs. It should be underlined, that spin AE currents might occur even in the case of equal populations of the valleys (in the absence of an external illumination).
The emergence of the spin current, together with the electric currents [Eqs.~\eqref{EqWarpingCurrent} and~\eqref{EqHallCurrent}], are the  quintessence of the valley AE effect.


\textit{{Conclusions.---}}
We have reported on the valley acoustoelectric effect and the valley acoustoelectric Hall effect in noncentrosymmetric materials exposed to surface acoustic waves. 
We calculated the electric current densities and compared their magnitudes and directions of propagation with the conventional diffusive current, suggesting a way to design topologically diverse patterns of electric current and the spin current. 


\paragraph*{ Acknowledgments.}
We have been supported by the Institute for Basic Science in Korea (Project No.~IBS-R024-D1) and the Russian Foundation for Basic Research (Project No. 19-42-540011).

\newpage

\begin{widetext}

\section*{Supplemental Material}

\subsection{I. The warping current}

Let us disregard the Berry phase effect in this Section and only consider the effect of warping.

The first-order corrections to the distribution function and the electron density,
%
$f_1(\textbf{r},t)=\left(f_1e^{i\textbf{k}\cdot\mathbf{r}-i\omega t}+f_1^*e^{-i\textbf{k}\cdot\mathbf{r}+i\omega t}\right)/2$ and
$n_1(\textbf{r},t)=
\left(n_1e^{i\textbf{k}\cdot\mathbf{r}-i\omega t}+n_1^*e^{-i\textbf{k}\cdot\mathbf{r}+i\omega t}\right)/2$,
%
should satisfy the Boltzmann transport equation
 \begin{gather}\label{eq3.1}
-i(\omega-\textbf{k}\cdot\mathbf{v}) f_1+e\Bigl(\textbf{E}+\textbf{E}^i\Bigr)\cdot\frac{\partial f_0}{\partial \textbf{p}}=-\frac{1}{\tau}\left(f_1-n_1\frac{\partial f_0}{\partial n}\right),
\end{gather}
which yields
\begin{gather}\label{eq4.1}
f_1=\frac{-e\tau(\textbf{E}+\textbf{E}^i)\cdot\partial_\textbf{p}f_0+n_1\partial_nf_0}
{1-i(\omega-\textbf{k}\cdot\mathbf{v})\tau},
\end{gather}
where $\textbf{v}=\textbf{p}/m+\partial_\textbf{p}w_\textbf{p}$. The general solution Eq.~(\ref{eq4.1}) might also contain the term $m\textbf{k}\cdot\dot{\textbf{U}}$, where $\textbf{U}$ is the displacement vector of the media~\cite{Kittel, Chaplik}. However, in the case of a transverse Bleustein-Gulyaev SAW, this term does not contribute to the effect we consider.

The induced electric field obeys the Maxwell's equation, which gives
\begin{eqnarray}\label{EqIndField}
\textbf{E}^i=-4\pi ie\textbf{k}n_1/(k\epsilon+k)
,
\end{eqnarray}
where $\epsilon$ is the dielectric constant of the substrate.
Combining the continuity equation with Eqs.~\eqref{eq4.1} and~\eqref{EqIndField}, and using the standard definition of the current density, we find the self-consistent solutions
\begin{gather}\label{eq7}
n_1=\frac{k_\alpha\sigma_{\alpha\beta}E_\beta}{e(\omega-\textbf{k}\cdot\mathbf{R})g(\textbf{k},\omega)},\\
\nonumber
g(\textbf{k},\omega)=
1+i\frac{4\pi }{\epsilon+1}\frac{k_\alpha\sigma_{\alpha\beta}k_\beta}{k(\omega-\textbf{k}\cdot\mathbf{R})},
\end{gather}
where $k=|\textbf{k}|$ and
\begin{gather}\label{eq8}
\sigma_{\alpha\beta}=e^2\tau\int\frac{d\textbf{p}}{(2\pi)^2}\frac{v_\alpha v_\beta}{1-i(\omega-\textbf{k}\cdot\mathbf{v})\tau}\left(-\frac{\partial f_0}{\partial\varepsilon_\textbf{p}}\right),\\\nonumber
\textbf{R}=\frac{\partial\mu}{\partial n}\int\frac{d\textbf{p}}{(2\pi)^2}\frac{\textbf{v}}{1-i(\omega-\textbf{k}\cdot\mathbf{v})\tau}\left(-\frac{\partial f_0}{\partial\varepsilon_\textbf{p}}\right)
\end{gather}
are the conductivity tensor of the 2DEG and the diffusion vector~\cite{Kittel}, respectively.

The stationary part of the second-order correction to the electron distribution function ($f_2$) satisfies the equation
\begin{eqnarray}\label{eq9.1}
&&\frac{e}{4}\left(\textbf{E}+\textbf{E}^i\right)\cdot\frac{\partial f_1^*}{\partial \textbf{p}}+
\frac{e}{4}\left(\textbf{E}^*+\textbf{E}^{i*}\right)\cdot\frac{\partial f_1 }{\partial\textbf{p}}
=-\frac{1}{\tau}\left(f_2-\bar{n}_2\frac{\partial f_0}{\partial n}-\frac{n_1n_1^*}{4}\frac{\partial^2 f_0}{\partial n^2}\right),
\end{eqnarray}
where $\bar{n}_2$ is the time-averaged second-order correction to the electron density.
It is easy to show that the second and third terms in the right hand side of Eq.~\eqref{eq9.1} do not contribute to the stationary electron current $\mathbf{j}=e\int d\textbf{p}f_2\textbf{v}/(2\pi)^2$.
Indeed, taking into account that
$$\textbf{v}\frac{\partial f_0}{\partial n}=\textbf{v}\frac{\partial \mu}{\partial n}\frac{\partial f_0}{\partial \mu}=
-\frac{\partial \mu}{\partial n}\textbf{v}\frac{\partial f_0}{\partial \varepsilon_\textbf{p}}=
-\frac{\partial \mu}{\partial n}\frac{\partial f_0}{\partial \textbf{p}},$$
the integral over the second term in~\eqref{eq9.1} gives zero since the last term here is an odd function of electron momentum. The same is true for the third term.
Thus, the current density yields
\begin{gather}\label{eq10.1}
j_\alpha=-\frac{e^2\tau}{2}\textmd{Re}\,\int\frac{d\textbf{p}}{(2\pi)^2}v_\alpha\left(E^*_\beta+E^{i*}_\beta\right)\frac{\partial f_1}{\partial p_\beta}.
\end{gather}
Since $\textbf{E}^i(\textbf{k},\omega)$ does not depend on the electron momentum $\textbf{p}$, we can simplify Eq.~\eqref{eq10.1} by integrating it by parts, which gives
\begin{gather}\label{eq11.1}
j_\alpha=\frac{e^2\tau}{2}\textmd{Re}\,\left(E^*_\beta+E^{i*}_\beta\right)\int\frac{d\textbf{p}}{(2\pi)^2}f_1\frac{\partial v_\alpha}{\partial p_\beta}.
\end{gather}
Now substituting Eq.~(\ref{eq4.1}) into Eq.~(\ref{eq11.1}), we find two terms of the current:
\begin{eqnarray}\label{eq12.1}
j^{(1)}_\alpha&=&\frac{e^3\tau^2}{2}\textmd{Re}\,\left(E^*_\beta+E^{i*}_\beta\right)\left(E_\gamma+E^{i}_\gamma\right)\\
\nonumber
&&\times\int\frac{d\textbf{p}}{(2\pi)^2}\frac{\partial v_\alpha}{\partial p_\beta}\frac{v_\gamma}{1-i(\omega-\textbf{k}\cdot\mathbf{v})\tau}\left(-\frac{\partial f_0}{\partial\varepsilon_\textbf{p}}\right)
\end{eqnarray}
and
%
%
%
\begin{gather}\label{eq13.1}
j^{(2)}_\alpha=\frac{e\tau}{2}\textmd{Re}\,\left(E^*_\beta+E^{i*}_\beta\right)\frac{k_\gamma\sigma_{\gamma\delta}}{\omega-\textbf{k}\cdot\mathbf{R}}
\left(E_\delta+E^{i}_\delta\right)
\times\\\nonumber
\times\frac{\partial\mu}{\partial n}\int\frac{d\textbf{p}}{(2\pi)^2}\frac{\partial v_\alpha}{\partial p_\beta}\frac{1}{1-i(\omega-\textbf{k}\cdot\mathbf{v})\tau}\left(-\frac{\partial f_0}{\partial\varepsilon_\textbf{p}}\right).
\end{gather}
%
Both $j^{(1)}$ and $j^{(2)}$ have acoustoelectric nature. However, their physical meaning is conceptually different.
The first contribution is a drift current, whereas the second one embodies the diffusive current.

We want to mention, that in the absence of the warping, a nonzero value of the drift current can be found by expanding $\textmd{Re}\,[1-i(\omega-\textbf{k}\cdot\mathbf{v})\tau]^{-1}$ in~\eqref{eq12.1} with respect to small $\omega\tau\ll\textbf{k}\cdot\mathbf{v}\tau\ll1$ and keeping the term $2\omega\tau\textbf{k}\cdot\mathbf{v}$ under the integral. 
In this case, the current is proportional to a small parameter $(\omega\tau) (kl)\ll1$, where $l=v_F\tau$ is the electron mean free path. 
In general, the drift current is much smaller than the diffusive term~\eqref{eq13.1} and we will neglect its contribution.

%
%
%
%


The valley AE current can be found by expanding expressions~(\ref{eq12.1}) and~(\ref{eq13.1}) over $w(\textbf{p})$. We encounter this term in the velocity $v_\alpha$ and the distribution function $f_0(\varepsilon_\textbf{p})$. Both of them we expand, only keeping the first--order corrections with respect to $w(\textbf{p})$. Thus, in the long-wavelength limit we neglect the term $\omega\tau-\textbf{k}\cdot\mathbf{v}\tau$ in the denominators of Eqs.~(\ref{eq12.1}) and~(\ref{eq13.1}). 
Consequently, the main contribution to the \textit{valley} AE current traces its origin to the drift term Eq.~(\ref{eq12.1}). 
Indeed, if we look at Eq.~(\ref{eq13.1}), the first--order terms (with respect to $w(\textbf{p})$) there vanish during the angle integration. 
A nonzero diffusion valley  current is only possible if we keep $(\omega\tau)(\textbf{k}\cdot\mathbf{v}\tau)$ in the denominator expansion. However, it has the smallness $kl\ll1$ ($\omega\tau\ll1$) in comparison with the drift term.

%
%
%
%
%
%

Further we find
\begin{gather}\label{eq15}
\frac{\partial v_\alpha}{\partial p_\beta}=\left(
                                             \begin{array}{cc}
                                               1/m+6\eta Cp_x & -6\eta Cp_y \\
                                               -6\eta Cp_y & 1/m-6\eta Cp_x \\
                                             \end{array}
                                           \right),
\end{gather}
and for an arbitrary direction of the SAW,
\begin{eqnarray}
\label{eq15}
j^{(W)}_x=j^{(1)}_x&=&3\eta Ce^3\tau^2n_e\Bigl[|E_x+E_x^i|^2-|E_y+E_y^i|^2\Bigr],\\\nonumber
j^{(W)}_y=j^{(1)}_y&=&-6\eta Ce^3\tau^2n_e\textmd{Re}\,(E_x^*+E_x^{i*})(E_y+E_y^i).
\end{eqnarray}
We immediately note, that the \textit{valley-dependent} acoustoelectric current does not depend on the direction of the wave vector of the acoustic wave, in contrast with the diffusive current Eq.~(3) from the main text.

Introducing the angle $\varphi$: $\textbf{k}=k(\cos\varphi,\sin\varphi)$ and hence $\textbf{E}=E_0(\cos\varphi,\sin\varphi)$, we find
\begin{eqnarray}\label{eq16}
j^{(W)}_x=j_0\cos2\varphi,
~~~~~
j^{(W)}_y=-j_0\sin2\varphi,
\end{eqnarray}
where $j_0$ for a degenerate electron gas at $T=0$ reads
\begin{eqnarray}\label{eq17}
j_0=3e\eta C(eE_0\tau)^2n_e\frac{1+(Dk^2/\omega)^2}{1+(kv^*/\omega+Dk^2/\omega)^2}.
\end{eqnarray}
Writing the SAW dispersion as $\omega=sk$, where $s$ is the SAW velocity, Eq.~(\ref{eq17}) transforms into $j_0=j_\eta F(\sigma)$, where
\begin{gather}\label{eq17.1}
j_\eta=3e\eta C(eE_0\tau)^2n_e,\\\nonumber
F(\sigma)=\frac{1+(\sigma/\sigma_*)^2(ka)^2}{1+(\sigma/\sigma_*)^2(1+ka)^2}. 
\end{gather}
Here $F(\sigma)$ is a screening factor, which  contains the total conductivity $\sigma$ summed over both the valleys; $\sigma_*=(\varepsilon+1)s/4\pi$ and $a=(\varepsilon+1)/(4me^2)$.

\subsection{II. The Hall current}

Let us disregard the effect of trigonal warping in this Section and only consider the effect of the Berry phase. (Some of the formulas in this Section will repeat similar formulas from Section~I for the convenience of reading.)

The group velocity describing the quasiclassical dynamics of a Bloch electron in the absence of an external magnetic field reads
\begin{gather}\label{EQ.1}
\dot{\textbf{r}}=\mathbf{v}-\dot{\textbf{p}}\times {\bf \Omega}_\textbf{p},
\end{gather} 
where $\mathbf{v}=\partial\varepsilon_{\textbf{p}}/\partial\textbf{p}$ and $\dot{\textbf{p}}=e\tilde{\textbf{E}}$, and $\tilde{\textbf{E}}(\textbf{r},t)=\tilde{\textbf{E}}e^{i\textbf{kr}-i\omega t}/2 +\textrm{c.c.}$ is the overall electric field, including the external and induced contributions.
%
The Berry curvature reads ${\bf \Omega}_\textbf{p}=\partial_\textbf{p}\times\langle u|i\partial_\textbf{p}|u\rangle$ and $|u\rangle$ is a periodic amplitude of the Bloch wave function. 
%
%
%
The Boltzmann transport equation for the electrons reads~\cite{Kittel, Chaplik}
\begin{gather}\label{EQ.2}
\frac{\partial f}{\partial t}+\dot{\textbf{p}}\cdot\frac{\partial f}{\partial \textbf{p}}+\dot{\textbf{r}}\cdot\frac{\partial f}{\partial \textbf{r}}=I\{f\},
\end{gather} 
where $f$ is the electron distribution function and $I\{f\}$ is the collision integral, discussed in the main text.

The current is
\begin{gather}\label{EQ.3}
\textbf{j}(\textbf{r},t)=e\int\frac{d\textbf{p}}{(2\pi)^2}\dot{\textbf{r}}f(\textbf{p},\textbf{r},t).
\end{gather}
Combining Eqs.~(\ref{EQ.1})-(\ref{EQ.3}) together we find
\begin{gather}\label{EQ.4}
\frac{\partial f}{\partial t}+e\tilde{\textbf{E}}\cdot\frac{\partial f}{\partial \textbf{p}}+\left(\frac{\partial\varepsilon_{\textbf{p}}}{\partial\textbf{p}}-e\tilde{\textbf{E}}\times {\bf \Omega}_\textbf{p}\right)\cdot\frac{\partial f}{\partial \textbf{r}}=I\{f\},\\
\label{EqCurrentGeneral}
\textbf{j}=e\int\frac{d\textbf{p}}{(2\pi)^2}\left(\frac{\partial\varepsilon_{\textbf{p}}}{\partial\textbf{p}}-e\tilde{\textbf{E}}\times {\bf \Omega}_\textbf{p}\right)f(\textbf{p},\textbf{r},t).
\end{gather} 
The calculation shows that the acoustoelectric current is the second-order response of the system to the piezoelectric field. 
To find it, we expand the distribution function in series: $f(\mathbf{p},\textbf{r},t)=f_0(\mathbf{p})+f_1(\mathbf{p},\textbf{r},t)+f_2(\mathbf{p},\textbf{r},t)+O(f_3)$, where $f_0$ is the equilibrium electron distribution. 
We also expand the density: $N(\textbf{r},t)=n+n_1(\textbf{r},t)+O(n_2)$, where $n=\int d\mathbf{p}f_0(\mathbf{p})/(2\pi)^2$ is the unperturbed electron density and $n_1$ is the first correction coming from the electron density fluctuations.
We further expand $ \bar{f}=f_0+(n_1+n_2+...)\partial_n f_0+(n_1+n_2+...)^2/2\partial^2 f_0/\partial n^2+O(f_0)$.
The first-order corrections read $f_1(\textbf{p},\textbf{r},t)=f_1(\textbf{p})e^{i\textbf{kr}-i\omega t}/2+\textrm{c.c.}$ and
$n_1(\textbf{r},t)=n_1e^{i\textbf{kr}-i\omega t}/2+\textrm{c.c.}$
[in what follows we will omit the explicit argument dependencies, using, e.g., $f_1\equiv f_1(\textbf{p})$].
They satisfy the Boltzmann equation 
 \begin{gather}\label{eq3}
-i(\omega-\textbf{k}\cdot\mathbf{v}) f_1+e\Bigl(\textbf{E}+\textbf{E}^i\Bigr)\frac{\partial f_0}{\partial \textbf{p}}=-\frac{1}{\tau}\left(f_1-n_1\frac{\partial f_0}{\partial n}\right),
\end{gather}
which yields
\begin{gather}\label{eq4}
f_1=\frac{-e\tau(\textbf{E}+\textbf{E}^i)\partial_\textbf{p}f_0+n_1\partial_nf_0}
{1-i(\omega-\textbf{k}\cdot\mathbf{v})\tau}.
\end{gather}

Using the continuity equation and summing up, we can express $n_1$ through the total electric field as
\begin{gather}\label{Eqn1FullField}
n_1=\frac{k_\alpha\tilde{\sigma}_{\alpha\beta}({E}_\beta+{E}_\beta^i)}{e(\omega-\textbf{k}\cdot\mathbf{R})},
\end{gather}
where $k=|\textbf{k}|$, $\tilde{\sigma}_{\alpha\beta}=\sigma_{\alpha\beta}-e^2\varepsilon_{\alpha\beta}\langle\Omega\rangle_z$ is the generalized conductivity tensor, with $\varepsilon_{\alpha\beta}$ the second-rank Levi-Civita permutation tensor ($\varepsilon_{xx}=\varepsilon_{yy}=0$, $\varepsilon_{xy}=-\varepsilon_{yx}=1$),  $\langle\Omega\rangle_z=n\Omega_0$ is $z$-component of the vector $\langle\mathbf{\Omega}\rangle=(2\pi)^{-2}\int d\mathbf{p}f_0(\mathbf{p})\mathbf{\Omega}_\mathbf{p}$, and
\begin{gather}\label{eq8}
\sigma_{\alpha\beta}=e^2\tau\int\frac{d\textbf{p}}{(2\pi)^2}\frac{v_\alpha v_\beta}{1-i(\omega-\textbf{k}\cdot\mathbf{v})\tau}\left(-\frac{\partial f_0}{\partial\varepsilon_\textbf{p}}\right),\\\nonumber
\textbf{R}=\frac{\partial\mu}{\partial n}\int\frac{d\textbf{p}}{(2\pi)^2}\frac{\textbf{v}}{1-i(\omega-\textbf{k}\cdot\mathbf{v})\tau}\left(-\frac{\partial f_0}{\partial\varepsilon_\textbf{p}}\right)
\end{gather}
are the tensor of conductivity and the diffusion vector~\cite{Kittel}, respectively. We can also express $n_1$ in terms of the piezoelectric field only:
\begin{gather}\label{eq7}
n_1=\frac{k_\alpha\tilde{\sigma}_{\alpha\beta}E_\beta}{e(\omega-\textbf{k}\cdot\mathbf{R})g(\textbf{k},\omega)},
\end{gather}
thus introducing the function
\begin{gather}\label{EqScreeningFunction}
g(\textbf{k},\omega)=1+i\frac{4\pi }{\epsilon+1}\frac{k_\alpha\sigma_{\alpha\beta}k_\beta}{k(\omega-\textbf{k}\cdot\mathbf{R})},
\end{gather}
which is responsible for the screening. Important to note, that the term containing the Berry curvature cancels out in the screening function.

Further we can find the stationary part of the second-order correction $f_2$, which satisfies the equation
\begin{eqnarray}\label{eq9}
&&\frac{e}{4}\left(\textbf{E}+\textbf{E}^i\right)\frac{\partial f_1^*}{\partial \textbf{p}}
+
\frac{e}{4}\left(\textbf{E}^*+\textbf{E}^{i*}\right)\frac{\partial f_1 }{\partial\textbf{p}}
=-\frac{1}{\tau}\left(
f_2
-\bar{n}_2\frac{\partial f_0}{\partial n}
-\frac{n_1n_1^*}{4}\frac{\partial^2 f_0}{\partial n^2}\right),
\end{eqnarray}
where $\bar{n}_2$ is the time-averaged $n_2$, but the two last terms in the equation above do not give any contributions (see the explanations below Eq.~\eqref{eq9.1} in Sec.~1).
%
%

From Eq.~\eqref{EqCurrentGeneral} we now can find the current density:
\begin{gather}\label{eq10}
j_\alpha
=-\frac{e^2}{2}
\textrm{Re}
\int\frac{d\textbf{p}}{(2\pi)^2}
\tau v_\alpha
\left(E^*_\beta+E^{i*}_\beta\right)\frac{\partial f_1}{\partial p_\beta}
\\
\nonumber
-
\Omega_z(\mathbf{p})
\varepsilon_{\alpha\beta}
\left(E^*_\beta+E^{i*}_\beta\right)f_1
.
\end{gather}
Since $\textbf{E}(\textbf{k},\omega)$ and $\textbf{E}^i(\textbf{k},\omega)$ are independent of the electron momentum $\textbf{p}$, we can extract them from the integral and integrate by parts the upper line in Eq.~\eqref{eq10}, yielding
\begin{gather}\label{eq11}
j_\alpha=\frac{e^2}{2}\textmd{Re}\,\left(E^*_\beta+E^{i*}_\beta\right)\int\frac{d\textbf{p}}{(2\pi)^2}f_1
\left(
\frac{\tau}{m}
\delta_{\alpha\beta}
-
\Omega_z(\mathbf{p})\varepsilon_{\alpha\beta}
\right),
\end{gather}
where we used $\partial v_\alpha/\partial p_\beta=\delta_{\alpha\beta}/m$.
This formula gives the general form of the square-in-field and linear-in-curvature electric current, including the conventional terms and the Hall effect.

The next step of the derivations is to substitute Eq.~(\ref{Eqn1FullField}) in Eq.~(\ref{eq4}) and then the result 
in Eq.~(\ref{eq11}). 
Before that, let us assume the long-wavelength limit ($\omega\tau\ll1$, $\mathbf{k}\cdot\mathbf{v}\tau\ll 1$), thus disregarding all the terms in the denominator in Eqs.~(\ref{eq4}) except for the unity.
Then the term $-e\tau(\textbf{E}+\textbf{E}^i)\partial_\textbf{p}f_0$ in the numerator of Eqs.~(\ref{eq4}) gives no contribution to the Hall current since it is proportional to $\mathbf{p}$ (coming from $\partial_\mathbf{p}f_0$) and all other terms contain $|\mathbf{p}|$ only, thus the integral $\int d\mathbf{p}$ vanishes.
We are only interested in the Hall contribution, in other words, in linear in Berry curvature terms (proportional to either $\Omega_z$ or the mean $\langle\Omega\rangle_z$), therefore we will omit all the other terms.
After all these assumptions, for the degenerate electron gas at $T=0$ (giving $\partial\mu/\partial n=\pi/m$ and $-\partial f_0/\partial \varepsilon_\mathbf{p}=\delta[\varepsilon_\mathbf{p}-\mu]$), we explicitly find
\begin{eqnarray}\label{eq12}
j^{(H)}_\alpha=-\frac{1}{2en}
\textmd{Re}\,
\left\{
\frac{\tilde{\sigma}_{\alpha\beta}(E^*_\beta+E_\beta^{i*})k_\gamma\tilde{\sigma}_{\gamma\eta}(E_\eta+E_\eta^i)}{\omega-\mathbf{k}\cdot\mathbf{R}}
\right\}.~~~~~~
\end{eqnarray}
Let us parametrize: $\mathbf{k}=k(\cos\phi,\sin\phi)$ and thus $\mathbf{E}=E_0(\cos\phi,\sin\phi)$. Then Eq.~\eqref{eq7} gives
\begin{gather}\label{Eqn1add}
n_1=\frac{k_x\tilde{\sigma}_{xx}E_x
+k_y\tilde{\sigma}_{yy}E_y+k_x\tilde{\sigma}_{xy}E_y
+k_y\tilde{\sigma}_{yx}E_x}{e(\omega-\textbf{k}\cdot\mathbf{R})g(\textbf{k},\omega)}
\end{gather}
or
\begin{gather}\label{Eqn1add2}
n_1=\frac{\sigma kE_0}{e(\omega-\textbf{k}\cdot\mathbf{R})g(\textbf{k},\omega)}
\end{gather}
since $\tilde\sigma_{xy}=-\tilde\sigma_{yx}$ and $\sigma_{\alpha\beta}=\sigma\delta_{\alpha\beta}$, where $\sigma=e^2n\tau/m$ is a static conductivity. 
We also note that $E_\beta+E_\beta^i=E_\beta/g(\mathbf{k},\omega)$, yielding
\begin{eqnarray}\label{EqCur3}
j^{(H)}_\alpha=-\frac{1}{2en}
\frac{1}{|g(\mathbf{k},\omega)|^2}
\textmd{Re}\,
\left\{
\frac{\tilde{\sigma}_{\alpha\beta}E^*_\beta k_\gamma\tilde{\sigma}_{\gamma\eta}E_\eta}{\omega-\mathbf{k}\cdot\mathbf{R}}
\right\}~~~~~~
\end{eqnarray}
or after some algebra
\begin{eqnarray}\label{EqCurMain}
j^{(H)}_\alpha=\frac{e\sigma}{2\omega}
\Omega_0
\frac{E_\beta k_\gamma E_\eta
(\delta_{\alpha\beta}\varepsilon_{\gamma\eta}
+
\varepsilon_{\alpha\beta}\delta_{\gamma\eta})
}
{1+(kv^*/\omega+Dk^2/\omega)^2}
,~~~~~~
\end{eqnarray}
where we kept only linear in $\Omega_0$ terms, assuming that the piezoelectric field amplitude is real, and denoting $D=v_F^2\tau/2$, $v_F$, and $v^*=4\pi\sigma/(\varepsilon+1)$ being the diffusion coefficient, Fermi velocity, and the charge spreading velocity, respectively. 
Finally we find
\begin{eqnarray}\label{EqHallCurrent}
\textbf{j}^{(H)}=\frac{e\sigma k}{2\omega}\frac{[\textbf{n}\times{\bf \Omega}_0]}{1+\left(\sigma/\sigma_*\right)^2(1+ka)^2}E_0^2,
\end{eqnarray}
where $\textbf{n}=\textbf{k}/k$.

%
%
%
%



%
%
\end{widetext}
%
%



\begin{thebibliography}{50}


\bibitem{RefRadisavljevic} B. Radisavljevic, A.~Radenovic, J.~Brivio, V.~Giacometti, A.~Kis, Single-layer MoS$_2$ transistors, Nature Nanotech. \textbf{6}(3),  147?50 (2011).

\bibitem{RefSundaram} R. S. Sundaram, M. Engel, A.~Lombardo, R.~Krupke, A.~C.~Ferrari, Ph.~Avouris, M.~Steiner, Electroluminescence in Single Layer MoS$_2$, Nano Lett. \textbf{13}(4), 1416 (2013).

\bibitem{xu2014spin} X. Xu, W. Yao, D. Xiao and T. F. Heinz, Spin and pseudospins in layered transition metal dichalcogenides, Nature Phys. \textbf{10}, 343 (2014).

\bibitem{Geim} K. S. Novoselov, A. K. Geim, S.~V.~Morozov, D.~Jiang, M.~I.~Katsnelson, I.~V.~Grigorieva, S.~V.~Dubonos, and A.~A.~Firsov, Two-dimensional gas of massless Dirac fermions in graphene,
Nature (London) \textbf{438}, 197-200 (2005).

\bibitem{xiao2012coupled} D Xiao, G.-B.~Liu, W.~Feng, X.~Xu, and W.~Yao, Coupled Spin and Valley Physics in Monolayers of MoS$_2$ and Other Group-VI Dichalcogenides,
Phys. Rev. Lett. \textbf{108}, 196802 (2012).

\bibitem{RefMak} K. F. Mak, K.~L.~McGill, J.~Park, P.~L.~McEuen, The valley Hall effect in MoS$_2$ transistors, Science
\textbf{344}(6191), 1489 (2014).

\bibitem{RefUbrig} N. Ubrig, S. Jo, M. Philippi, D.~Costanzo, H.~Berger, A.~B.~Kuzmenko, and A.~F.~Morpurgo, Microscopic origin of the valley Hall effect in transition metal dichalcogenides revealed by wavelength-dependent mapping, Nano Lett. \textbf{17}, 5719 (2017).

\bibitem{OurNJP} V. M. Kovalev, W.-K. Tse, M.~V.~Fistul, and I.~G.~Savenko, Valley Hall transport of photon-dressed quasiparticles in two-dimensional Dirac semiconductors, New~J.~Phys. \textbf{20}, 083007 (2018).

\bibitem{ShelykhGap} O. V. Kibis, K. Dini, I.~V.~Iorsh, and I.~A.~Shelykh, All-optical band engineering of gapped Dirac materials, Phys. Rev. B \textbf{95}, 125401 (2017).

\bibitem{Machlin} J. Tuorila, M. Silveri, M.~Sillanp\"a\"a, E.~Thuneberg, Yu.~Makhlin, and P.~Hakonen, Stark effect and generalized Bloch-Siegert shift in a strongly driven two-level system, Phys. Rev. Lett. \textbf{105}, 257003 (2010).

\bibitem{BlochOsc} L. Allen and J. H. Eberly, \textit{Optical Resonance and Two-Level Atoms}
(Dover Publications, 1987).

\bibitem{KibisPRL} O. Kibis, Dissipationless Electron Transport in Photon-Dressed Nanostructures, Phys. Rev. Lett. \textbf{107}, 106802 (2011).

\bibitem{Wieck} A. D. Wieck, H. Sigg, and K.~Ploog, Observation of resonant photon drag in a two-dimensional electron gas, Phys. Rev. Lett. \textbf{64}, 463 (1990).

\bibitem{Glazov} M. M. Glazov and S. D. Ganichev, High frequency electric field induced nonlinear effects in graphene, Phys. Rep.~\textbf{535}, 101 (2014).

\bibitem{Entin} M. V. Entin, L. I. Magarill, and D.~L.~Shepelyansky, Theory of resonant photon drag in monolayer graphene, Phys. Rev. B \textbf{81}, 165441 (2010).

\bibitem{RefRPQ} M.~V.~Boev, V.~M.~Kovalev, I.~G.~Savenko, Resonant photon drag of dipolar excitons, JETP Lett. \textbf{107}, 763 (2018);
V.~M.~Kovalev, M.~V.~Boev, I.~G.~Savenko, Proposal for frequency-selective photodetector based on the resonant photon drag effect in a condensate of indirect excitons, Phys. Rev. B \textbf{98}, 041304(R) (2018).

\bibitem{RefBasov} Zh.~Sun, D.~N.~Basov, and M.~M.~Fogler, Third-order optical conductivity of an electron fluid, Phys. Rev. B \textbf{97}, 075432 (2018).

\bibitem{OurRecPRB} L. I. Magarill, M. V. Entin and V.~M.~Kovalev, arXiv:1811.01187 (2018); V. M. Kovalev and I. G. Savenko, Photogalvanic currents in dynamically gapped transition metal dichalcogenide monolayers, Phys. Rev. B \textbf{99}, 075405 (2019).


\bibitem{belinicher}  V. I. Belinicher and B.~I.~Sturman,  The photogalvanic effect in media lacking a center of symmetry, Sov. Phys. Usp. \textbf{23}, 199 (1980).

\bibitem{Shan} W.-Y.~Shan, J. Zhou, and Di~Xiao, Optical generation and detection of pure valley current in monolayer transition-metal dichalcogenides, Phys. Rev. B \textbf{91}, 035402 (2015).

\bibitem{MEGT} L. E. Golub, S. A. Tarasenko, M.~V.~Entin and L.~I.~Magarill, Valley separation in graphene by polarized light, Phys. Rev. B \textbf{84}, 195408 (2011).

\bibitem{Hongyi} H. Yu, Y. Wu, G.-B.~Liu, X.~Xu, and W.~Yao, Nonlinear Valley and Spin Currents from Fermi Pocket Anisotropy in 2D Crystals, Phys. Rev. Lett. \textbf{113}, 156603 (2014).

\bibitem{ZhangScience} Y. J. Zhang, T. Oka, R. Suzuki, J.~T.~Ye, and Y.~Iwasa, Electrically switchable chiral light-emitting transistor, Science \textbf{344}, 725 (2014).

\bibitem{GT} L. E. Golub and S. A. Tarasenko, Valley polarization induced second harmonic generation in graphene, Phys. Rev. B \textbf{90}, 201402(R) (2014).


\bibitem{Portnoi1}  R. R. Hartmann and M. E. Portnoi, Optoelectronic Properties of
Carbon-based Nanostructures: Steering electrons in graphene
by electromagnetic fields (LAP LAMBERT Academic Publishing,
Saarbrucken, 2011).

\bibitem{RefChang} M.-C.~Chang and Q. Niu, Berry phase, hyperorbits, and the Hofstadter spectrum: Semiclassical dynamics in magnetic Bloch bands, Phys. Rev. B \textbf{53}, 7010 (1996).

\bibitem{RefBerry} M. V. Berry, Quantal phase factors accompanying adiabatic changes, Proc. R. Soc. Lond. A Math. Phys. Sci. \textbf{392}, 45 (1984).

\bibitem{RefNagaosa} N.~Nagaosa, J.~Sinova, S.~Onoda, A.~H.~MacDonald, and N.-P.~Ong, Anomalous Hall effect, Rev. Mod. Phys. \textbf{82}, 1539 (2010).

\bibitem{RefThouless} D. J. Thouless, M. Kohmoto, M.~P.~Nightingale, M.~den~Nijs, Quantized Hall Conductance in a Two-Dimensional Periodic Potential,
Phys. Rev. Lett. \textbf{49}, 405 (1982).

\bibitem{RefHasan} M. Z. Hasan, C. L. Kane, Topological insulators, Rev. Mod. Phys. \textbf{82}, 3045 (2010).

\bibitem{RefNiu} Q. Niu and D. J. Thouless, Quantised adiabatic charge transport in the presence of substrate disorder and many-body interaction, J. Phys. A \textbf{17}, 2453 (1984).

\bibitem{RefXiao} D.~Xiao, Y.~Yao,Zh.~Fang, and Q.~Niu, Berry-Phase Effect in Anomalous Thermoelectric Transport, Phys. Rev. Lett. \textbf{97}, 026603 (2006).

\bibitem{RefXiaoRMP} D.~Xiao, M.-C.~Chang and Q. Niu, Berry phase effects on electronic properties, Rev. Mod. Phys. \textbf{82}, 1959 (2010).





\bibitem{RefZhang} Y. Zhang, Y. W. Tan, H. L. Stormer, P. Kim, Experimental observation of the quantum Hall effect and Berry's phase in graphene, Nature \textbf{438}, 201 (2005).

\bibitem{RefYou} J.-S.~You, S.~Fang, S.-Y.~Xu, E.~Kaxiras, and T.~Low, Berry curvature dipole current in the transition metal dichalcogenides family, Phys. Rev. B \textbf{98}, 121109(R) (2018).

%
%
%

\bibitem{RefWixforth} A. Wixforth, J. Scriba, M.~Wassermeier, J.~P.~Kotthaus, G.~Weimann, and W.~Schlapp, Surface acoustic waves on GaAs/Al$_x$Ga$_{1-x}$As heterostructures,
Phys. Rev. B \textbf{40}, 7874 (1989).

\bibitem{RefWillet} R. L. Willet, M. A. Paalanen, R.~R.~Ruel, K.~W.~West, L.~N.~Pfeiffer, and B.~J.~Bishop,
Anomalous sound propagation at $\nu=1/2$ in a 2D electron gas: Observation of a spontaneously broken translational symmetry?
Phys. Rev. Lett. \textbf{65}, 112 (1990).

\bibitem{graphene1} S. H. Zhang and W. Xu, Absorption of surface acoustic waves by graphene, AIP Advances \textbf{1}, 022146 (2011).

\bibitem{graphene2} V. Miseikis, J. E. Cunningham, K. Saeed, R. O'Rorke, and A. G. Davies, Acoustically induced current flow in graphene, Appl. Phys. Lett. \textbf{100}, 133105 (2012).

\bibitem{top1} V. Parente, A. Tagliacozzo, F.~von~Oppen, and F.~Guinea, Electron-phonon interaction on the surface of a three-dimensional topological insulator,  Phys. Rev. B \textbf{88}, 075432 (2013).

\bibitem{top2} L. L. Li and W. Xu, Absorption of surface acoustic waves by topological insulator thin films, Appl. Phys. Lett. \textbf{105}, 063503 (2014).

\bibitem{reviewkovalev1} E.~G.~Batyev, V.~M.~Kovalev, A.~V.~Chaplik, Response of a Bose-Einstein condensate of dipole excitons to static and dynamic perturbations, JETP Lett. \textbf{99}(9), 540 (2014).

\bibitem{reviewkovalev2} M. V. Boev, A. V. Chaplik , V.~M.~Kovalev, Interaction of Rayleigh waves with 2D dipolar exciton gas: Impact of Bose-Einstein condensation, J. Phys. D: Appl. Phys. \textbf{50}(48), 484002 (2017).

\bibitem{dragkovalev1} V.~M.~Kovalev, A.~V.~Chaplik, Effect of exciton dragging by a surface acoustic wave,	JETP Lett. \textbf{101}(3), 177 (2015);
Acousto-exciton interaction in a gas of 2D indirect dipolar excitons in the presence of disorder, JETP \textbf{122}(3), 499 (2016).


\bibitem{RefEsslingen} A. Esslingen, A. Wixforth, R. W. Winkler, J. P. Kotthaus, H. Nickel, W. Schlapp, and R. Losch, Solid State Commun. \textbf{84}, 939 (1992).

\bibitem{Parmenter} R. H. Parmenter, The Acousto-Electric Effect, 
Phys. Rev. \textbf{89}, 990 (1953).


%
%
%
%
%
%
%
%
%
%




















%
%
%
%
%

\bibitem{RefFalkoPar} A.~Kormanyos, G.~Burkard, M.~Gmitra, J.~Fabian, V.~Zolyomi, N.~D.~Drummond, and V.~Fal'ko, \textbf{k}\textbf{p} theory for two-dimensional transition metal dichalcogenide semiconductors, 2D Mater. \textbf{2}, 049501 (2015).

\bibitem{Chaplik} M.~V.~Krasheninnikov and A.~V.~Chaplik, Plasma-acoustic waves on the surface of a piezoelectric crystal, JETP~\textbf{48}(5), 960 (1978).

\bibitem{Kittel} C.~Kittel, Quantum theory of solid states (Wiley, 2004).











































\bibitem{SM} see the Supplemental Materials for the details of the derivations of the trigonal warping and Hall electric current densities.

\bibitem{Falko1993} V. I. Falko, S. V. Meshkov, and S.~V.~Iordanskii, Acoustoelectric drag effect in the two-dimensional electron gas at strong magnetic field, Phys. Rev. B \textbf{47}, 9910 (1993).



\end{thebibliography}
\end{document}